\documentclass[11pt]{article}
\usepackage[letterpaper,margin=0.92in]{geometry}
\usepackage[T1]{fontenc}
\usepackage{lmodern}
\usepackage{algorithm}
\usepackage{algpseudocode}
\usepackage{amsmath,amssymb,amsthm,mathtools}
\usepackage{microtype}
\usepackage{xcolor}
\usepackage{tikz}
\usepackage{booktabs,tabularx,colortbl}
\usetikzlibrary{arrows.meta,calc,positioning,decorations.pathreplacing}
\usepackage[section]{placeins}
\usepackage[hidelinks]{hyperref}
\hypersetup{pdftitle={Metric Data Placement with Installation Costs and Integral Site Capacities},pdfauthor={}}
\newtheorem{theorem}{Theorem}[section]
\newtheorem{lemma}[theorem]{Lemma}
\newtheorem{remark}[theorem]{Remark}
\newtheorem{definition}[theorem]{Definition}
\newcommand{\R}{\mathcal R}
\newcommand{\calP}{\mathcal P}
\newcommand{\OPT}{\operatorname{OPT}}
\newcommand{\pos}[1]{\left(#1\right)_+}
\definecolor{navy}{RGB}{36,75,128}
\definecolor{teal}{RGB}{28,121,110}
\definecolor{gold}{RGB}{180,115,34}
\tikzset{
    flowball/.style={draw=navy,fill=navy!7,rounded corners=2pt,
                      inner sep=3.2pt,minimum height=5.5mm},
    flowballtwo/.style={draw=teal,fill=teal!7,rounded corners=2pt,
                      inner sep=3.2pt,minimum height=5.5mm},
    flowgate/.style={draw=black!75,fill=black!5,circle,inner sep=2.6pt},
    flowarrow/.style={-{Latex[length=1.65mm]},line width=0.55pt},
    primary/.style={-{Latex[length=1.7mm]},line width=0.8pt,draw=navy},
    secondary/.style={-{Latex[length=1.7mm]},line width=0.8pt,draw=teal}
}
\title{An Improved $(1\!+\!2e)$-Approximation Algorithm for Metric Data Placement}
\author{
S. Rasoul Etesami\\
\small Department of Industrial and Systems Engineering\\
\small Coordinated Science Laboratory\\
\small University of Illinois Urbana-Champaign, Urbana, IL, USA\\ 
\small \texttt{etesami1@illinois.edu}
}

\date{}
\begin{document}
\maketitle
\begin{abstract}
We study the metric data placement problem with site-capacity
constraints and fixed installation costs, aiming to minimize
the total cost of placing data (resources) and accessing their nearest
copies. We develop a deterministic polynomial-time approximation
algorithm based on a natural linear programming relaxation of
the problem. Starting from an optimal fractional solution, we
construct multiscale metric-ball partitions independently for
each resource and connect overlapping balls at consecutive
levels to form a flow network. The resource layers are then
coupled through shared site-capacity arcs, allowing us to
obtain an integral placement by minimizing a surrogate cost
over a single min-cost circulation. By optimizing the choice
of ball radii, we establish an approximation factor of
$1+2e\approx 6.44$. To our knowledge, this is the best
approximation guarantee for the metric data placement problem
since its introduction by Baev and Rajaraman in 2001
\cite{BaevRajaraman2001}.
\par\vspace{0.5em}
\end{abstract}

\section{Introduction}\label{sec:intro}

The efficient placement of data (resources) across distributed networks is a fundamental problem in network optimization, with applications in content delivery networks, web caching, cloud computing, and distributed storage systems. In these applications, a collection of servers stores and provides access to different resources, with each server subject to limited storage capacity. Since storing every resource at every server is generally infeasible, resources must be replicated selectively, and requests for locally unavailable resources must be served by other servers in the network. The resulting performance depends on both resource placement decisions and the communication costs incurred when accessing remote resources. An efficient placement must therefore balance the cost of storing resources against the cost of accessing their copies, subject to the storage capacity constraints of individual servers.

This trade-off gives rise to the \emph{data placement
problem}, introduced by Baev and Rajaraman
\cite{BaevRajaraman2001}, in which a collection of
resources must be distributed among capacity-constrained
sites to minimize the total access cost. In its more general
formulation, each site has a prescribed capacity, installing
a resource at a site incurs a fixed cost, and each client
may have a different demand for each resource. The access
cost is determined by the distance to the nearest installed
copy of the requested resource. Although the objective is
separable across resources, the site-capacity constraints
couple their placement decisions: installing one resource
at a site reduces the capacity available for other resources.
Consequently, even under metric access costs, the problem
cannot be decomposed into independent resource placement
problems without accounting for these capacity constraints.

The data placement problem has been studied from several
perspectives, including facility location, combinatorial
approximation algorithms, game theory, and distributed
resource allocation. We briefly review the most relevant
literature in three categories.

\subsection{Related Work}

\paragraph{Facility location and related clustering problems.}
A closely related class of problems is the classical
facility location problem, which seeks to open facilities
and assign clients to them while minimizing the total
facility opening and connection costs. In particular,
the metric uncapacitated facility location problem
has been extensively studied using linear programming,
primal-dual methods, local search, and greedy algorithms
\cite{ShmoysTardosAardal1997,JainVazirani2001,
JainMahdianEtAl2003,MahdianYeZhang2006}.
The best-known polynomial-time approximation factor
for this problem is $1.488$, achieved by Li
\cite{Li2013}, while obtaining an approximation factor
better than approximately $1.463$ is NP-hard
\cite{GuhaKhuller1999}. Related problems, including
$k$-median and its variants, have also inspired
extensive research on metric clustering and LP rounding
\cite{makarychev2021approximation,CharikarGuhaTardosShmoys2002}.

The connection to data placement is particularly
transparent when there is only one resource: opening
a facility corresponds to installing the resource at
a site, while connecting a client to a facility corresponds
to accessing an installed copy. With multiple resources,
however, each resource induces a separate facility
location problem, while the site-capacity constraints
couple the placement decisions across these problems.
In particular, relaxing the site-capacity constraints
using Lagrangian multipliers decomposes the data placement
problem into a collection of uncapacitated facility location
problems with modified installation costs
\cite{Etesami2024}. This decomposition provides insight
into the problem's structure, but the resulting independent
placements need not satisfy the original capacity constraints.
More general facility location models, such as matroid
median \cite{KrishnaswamyEtAl2011,Swamy2016} and knapsack
median \cite{KrishnaswamyLiSandeep2018}, offer additional
connections and algorithmic tools for handling constraints
on facility openings.

\vspace{-0.3cm}
\paragraph{Game-theoretic models of data placement.}
Another important research direction concerns distributed
systems in which individual agents control their own
storage decisions. In such settings, servers may belong
to different stakeholders who choose which resources
to store based on their own interests rather than
a centralized objective. This perspective has
motivated research on selfish caching and replication
games \cite{ChunEtAl2004,GopalakrishnanEtAl2012,
LaoutarisEtAl2006,EtesamiBasar2017}. These studies
examine the existence and computation of Nash equilibria,
the efficiency of decentralized storage decisions
as measured by the price of anarchy, and the impact
of capacity constraints on strategic resource allocation.

The work \cite{Etesami2020} studied
both the centralized optimization problem and its
game-theoretic counterpart, establishing complexity
results, approximation guarantees, and connections
between Nash equilibria and locally optimal
solutions of weighted Max-$k$-Cut problems.
More recently, Etesami \cite{Etesami2024}
considered data placement with general \emph{non-metric}
access costs and developed distributed algorithms
based on potential games, Glauber dynamics, and
auction mechanisms to obtain optimal data placements
with theoretical performance guarantees. These methods
provide alternative approaches to coordinating resource
allocation when centralized optimization is undesirable
or when communication costs do not satisfy the metric
property. In contrast, the present work focuses on
centralized optimization under metric access costs,
with the goal of improving the worst-case
approximation guarantee.

\vspace{-0.3cm}
\paragraph{Approximation algorithms for data placement.}

The approximation complexity of metric data placement
was first investigated by Baev and Rajaraman
\cite{BaevRajaraman2001}, who developed a
$20.5$-approximation algorithm based on linear
programming and clustering. Their work initiated
the study of constant-factor approximation algorithms
for placing multiple resources under hard
storage constraints. Guha and Munagala
\cite{GuhaMunagala2002} subsequently studied a
related model in which caches are subject to additional
constraints on the number of clients they can serve,
obtaining a constant-factor approximation guarantee
with bounded capacity augmentation. For the standard
metric data placement problem, Baev, Rajaraman,
and Swamy \cite{BaevRajaramanSwamy2008} improved
the approximation factor to $10$ while accommodating
heterogeneous demands and installation costs.
Their approach combines LP-based clustering with
network-flow rounding to obtain feasible integral
placements.

A significant improvement was achieved by Swamy
\cite{Swamy2016}, who observed that metric data
placement can be formulated as a special case of
matroid median. By developing an LP-rounding
algorithm for matroid median, Swamy established
an $8$-approximation guarantee for data placement,
including resource-dependent installation costs.
The approach combines clustering and half-integral
solutions with the integrality properties of
matroid-related polyhedra. Subsequently,
Krishnaswamy, Li, and Sandeep
\cite{KrishnaswamyLiSandeep2018} developed
a general iterative-rounding framework that improved
the approximation factor for matroid median to
$7.081$. However, this improvement applies to
the connection-cost formulation without facility-opening
costs and therefore yields the same guarantee only
for the zero-installation-cost variant of data placement.
It does not directly establish a $7.081$-approximation
for the general data placement problem with arbitrary
installation costs. Further improvements have been achieved for
related constrained clustering problems. For example,
Gupta, Moseley, and Zhou \cite{GuptaMoseleyZhou2025}
developed a structural iterative-rounding framework
that achieves a $(6.387+\epsilon)$-approximation
for knapsack median. Although this result demonstrates
the effectiveness of more refined rounding techniques,
the approximation guarantee does not extend
to the multiple-resource data placement problem with
coupled site-capacity constraints and installation costs.

A different direction was explored by Etesami
\cite{Etesami2020}, who developed a simple combinatorial
greedy algorithm achieving a $3$-approximation for the
homogeneous metric resource allocation problem, in which demand weights may
vary across sites but are identical across resources
at each site.
The algorithm also provides a $3$-competitive guarantee
in the online setting. Although this result substantially
improves the approximation factor under homogeneity,
the heterogeneous setting with resource-dependent
demands and installation costs remains more challenging. Table~\ref{tab:approx-comparison}
summarizes the main approximation guarantees
relevant to the present work.

\begin{table}[t]
\centering
\caption{Approximation guarantees for metric data placement}
\label{tab:approx-comparison}
\vspace{2pt}
\footnotesize
\setlength{\tabcolsep}{3pt}
\renewcommand{\arraystretch}{1.15}
\begin{tabularx}{\linewidth}{
@{}
>{\raggedright\arraybackslash}p{0.23\linewidth}
>{\centering\arraybackslash}p{0.17\linewidth}
>{\raggedright\arraybackslash}X
@{}}
\toprule
\textbf{Reference} &
\textbf{Ratio} &
\textbf{Setting and technique} \\
\midrule
Baev and Rajaraman
\cite{BaevRajaraman2001}
& $20.5$
& Metric placement; LP clustering and rounding. \\

Baev et al.
\cite{BaevRajaramanSwamy2008}
& $10$
& General model; LP and network-flow rounding. \\

Swamy \cite{Swamy2016}
& $8$
& General model; reduction to matroid median. \\

Krishnaswamy et al.
\cite{KrishnaswamyLiSandeep2018}
& $7.081$
& No installation costs; iterative rounding. \\

Etesami \cite{Etesami2020}
& $3$
& Homogeneous setting; combinatorial online greedy algorithm. \\

\midrule
\rowcolor{black!5}
\textbf{This work}
& $\boldsymbol{1\!+\!2e\approx 6.44}$
& \textbf{General model; multiscale clustering and min-cost flow.} \\
\bottomrule
\end{tabularx}
\vspace{-2mm}
\end{table}

\subsection{Our Approach and Contributions}

In this paper, we develop a deterministic
polynomial-time $(1+2e)$-approximation algorithm
for the metric data placement problem with
heterogeneous demands, arbitrary nonnegative
installation costs, and integral site capacities.
Our algorithm improves upon the previous
$8$-approximation guarantee for the general
model and, to the best of our knowledge, provides
the strongest approximation guarantee for this
problem since its introduction in 2001.
Beyond the improved approximation factor, a key
feature of our approach is its simple structure:
we solve a natural LP relaxation, construct
multiscale clusters separately for each resource,
and combine the resulting resource layers into
a single network flow problem that enforces
the original site-capacity constraints.
Our approach requires neither iterative rounding
nor joint clustering across resources. 

The starting point is an alternative representation
of the fractional access cost that makes the
underlying geometry explicit. Rather than working
directly with client-to-server assignment variables,
we project out these variables and express the fractional
access cost as
\[
C(y)=
\sum_{i,\ell}w_{i\ell}
\int_0^D
\bigl(1-y_\ell(B(i,r))\bigr)_+\,dr,
\]
where $y_\ell(B(i,r))$ denotes the total fractional
mass of resource $\ell$ inside the metric ball
$B(i,r)$. This continuous representation admits a natural
interpretation: the access cost is obtained by
integrating the missing resource mass over
increasing distance thresholds. It also suggests
a direct rounding principle: preserving sufficient
resource mass within appropriately chosen metric
balls at different distance scales. Thus, the
geometric objects used by our algorithm arise
naturally from the objective function itself,
rather than from an auxiliary clustering or
charging construction.

We implement this idea through a multiscale
clustering procedure. For each resource and
each prescribed radius, we greedily select
pairwise disjoint metric balls based on their
fractional resource masses. The resulting
partitions are constructed independently for
each resource and across
levels. Nevertheless, the greedy selection
ensures that every candidate ball intersects
a selected ball with at least as much fractional
resource mass. We then construct a directed
network whose nodes represent the selected balls
and whose arcs connect overlapping balls at
consecutive levels. Each resource has its own
collection of ball layers, while all resources
share the capacity arcs associated with the
physical sites. Thus, a single network integrates
the geometric information obtained independently
for each resource while enforcing the global
feasibility constraints of the original
placement problem.

The fractional LP solution induces a feasible
reference circulation on this network. To round
this solution, we introduce a surrogate objective
based on the amount of missing flow through the
selected ball nodes. The same geometric
interpretation that motivates the fractional
objective allows us to bound the true access
cost of an integral circulation by this surrogate,
which incorporates a correction for the additional
distances introduced by clustering.
Crucially, the surrogate is a separable,
piecewise-linear convex function of the network
throughputs and can be minimized using a
min-cost circulation formulation with integral
capacities. Consequently, an optimal circulation
can be chosen to be integral, and the induced
placement automatically satisfies all
site-capacity constraints.

The approximation analysis then reduces to
controlling the geometric correction terms.
By choosing the ball radii from a suitably
shifted geometric sequence and optimizing
over the shift, we show that the contribution
of these terms is bounded by $2e$ times the
fractional access cost. An important feature of this guarantee is
that the installation cost retains a coefficient
of one, while the approximation loss arises
solely from bounding the access cost.
Moreover, the entire rounding procedure
requires only a single globally coupled
min-cost circulation, without constructing
intermediate half-integral placements or
performing separate rounding procedures
for different resources.

Overall, our approach establishes a direct
connection between the continuous geometric
structure of the fractional access cost and
the integrality of network flows. Its main
advantage lies in combining resource-wise
greedy clustering with a single network flow to handle the coupling
among resources, yielding a relatively simple
approximation algorithm with an improved
approximation guarantee.

\subsection{Organization}

The rest of the paper is organized as follows.
Section~2 formulates the metric data placement
problem and derives its fractional relaxation.
Section~\ref{sec:balls} presents the multiscale
greedy ball partitions and constructs the
globally coupled circulation network.
Section~\ref{sec:rounding} develops the
surrogate flow objective and establishes
the rounding guarantee.
Section~\ref{sec:scales} then optimizes
the choice of ball radii and proves the
$(1+2e)$-approximation guarantee.
Finally, Section~\ref{sec:conclusion} concludes
the paper and discusses directions for future research.

\section{Problem Formulation}

In this section, we formally introduce the data placement problem
\cite{BaevRajaraman2001,BaevRajaramanSwamy2008}
and develop a natural fractional relaxation that serves as
the foundation for our approximation algorithm.

Let $(V,d)$ be a finite metric space, where $V$ is a set of
$n\ge 2$ sites that also serve as clients, and let $\mathcal R$
be a finite set of $m$ data resources. Each site $j\in V$ has an
integer capacity $b_j\ge 0$, representing the maximum number
of distinct unit-size resources that can be stored at that site.
Installing resource $\ell\in\mathcal R$ at site $j$ incurs a
fixed cost $f_{j\ell}\ge 0$. Furthermore, each client $i\in V$ has a
nonnegative demand weight $w_{i\ell}$ for resource $\ell$,
and accessing a copy of resource $\ell$ stored at site $j$
incurs a cost proportional to the distance $d(i,j)$.
We assume that all input parameters are rational numbers.

A \emph{feasible} placement is a collection
$S=(S_\ell)_{\ell\in\mathcal R}$, where
$S_\ell\subseteq V$ denotes the nonempty set of sites
storing resource $\ell$, subject to the capacity constraints
\[
    \sum_{\ell\in\mathcal R}
    \mathbf 1\{j\in S_\ell\}\le b_j,
    \qquad j\in V.
\]
Each client accesses its nearest installed copy of each
resource. Consequently, the total access cost is
\[
    C(S):=\sum_{i\in V}\sum_{\ell\in\mathcal R}
    w_{i\ell}d(i,S_\ell),
\]
where
\[
    d(i,S_\ell):=\min_{j\in S_\ell}d(i,j)
\]
is the minimum distance from client $i$ to a site storing resource $\ell$. Moreover, the total installation cost is given by
\[
    F(S):=\sum_{\ell\in\mathcal R}
    \sum_{j\in S_\ell}f_{j\ell}.
\]
The objective is to find a feasible placement minimizing
the sum of access and installation costs, i.e.,
\begin{align}\label{eq:combinatorial-cost}
\OPT=\min_{S\text{ feasible}}\{C(S)+F(S)\}.
\end{align}

If a resource is not installed at any site, we define its access cost to be infinite, thereby ensuring that every resource is installed at least once in any feasible solution with finite cost. Furthermore, we assume that the total capacity of the system is at least the number of resources, i.e., $\sum_{j\in V} b_j \ge m$,
which guarantees the existence of a feasible placement and, consequently, that the optimal value $\OPT$ is finite.

\subsection{An Integer Programming Formulation}

The above data placement problem can be equivalently written as an integer program (IP) \cite{BaevRajaramanSwamy2008}. For every $i,j\in V$ and $\ell\in\mathcal R$, define the binary variables
\[
y_{j\ell}=
\begin{cases}
1, & \text{if resource $\ell$ is installed at site $j$},\\
0, & \text{otherwise},
\end{cases}
\]
and
\[
x_{ij\ell}=
\begin{cases}
1, & \text{if client $i$ accesses resource $\ell$
     at site $j$},\\
0, & \text{otherwise}.
\end{cases}
\]
Then, it is easy to see that the data placement problem \eqref{eq:combinatorial-cost} admits the following IP
formulation:
\begin{equation}\label{eq:IP}
\begin{aligned}
\OPT=\min_{x,y}\quad&
\sum_{i,j\in V}\sum_{\ell\in\mathcal R}
w_{i\ell}d(i,j)x_{ij\ell}
+\sum_{j\in V}\sum_{\ell\in\mathcal R}
f_{j\ell}y_{j\ell}\\
\text{subject to}\quad&
\sum_{j\in V}x_{ij\ell}=1,
&&\forall i\in V,\ \ell\in\mathcal R,\\
&
x_{ij\ell}\le y_{j\ell},
&&\forall i,j\in V,\ \ell\in\mathcal R,\\
&
\sum_{\ell\in\mathcal R}y_{j\ell}\le b_j,
&&\forall j\in V,\\
&
y_{j\ell},\ x_{ij\ell}\in\{0,1\},
&&\forall i,j\in V,\ \ell\in\mathcal R.
\end{aligned}
\end{equation}

\subsection{Fractional Relaxation}

A key observation from the integer program~\eqref{eq:IP} is that the placement variables $y$ determine the optimal assignment variables $x$. Indeed, once the resource placements are fixed, each client $i$ can be assigned to the nearest site storing resource $\ell$, thereby minimizing its access cost. Consequently, the assignment variables $x$ can be eliminated, allowing us to reformulate the problem entirely in terms of the placement variables $y$. By subsequently relaxing the integrality constraints and allowing $y_{j\ell}\in[0,1]$, we obtain a fractional relaxation whose optimal value provides a lower bound on $\OPT$. 

To formulate this relaxation, let $D=\max_{i,j\in V}d(i,j)$, $\delta=\min_{\substack{i\neq j\in V}}d(i,j)>0$, and denote by
\[
    B(i,r)=\{j\in V:d(i,j)\le r\}
\]
the closed metric ball of radius $r\ge 0$ centered at
site $i\in V$. The fractional feasible region is
\begin{equation}\label{eq:feasible-y}
Y=\bigg\{y\in [0,1]^{nm}:\qquad
 \sum_{\ell\in\R} y_{j\ell}\le b_j \ \ \forall j\in V,\qquad
 \sum_{j\in V}y_{j\ell}\ge1\ \ \forall \ell\in\R\bigg\},
\end{equation}
where the last set of constraints in \eqref{eq:feasible-y} is obtained by summing the second set of constraints in \eqref{eq:IP} over $j \in V$ and applying the first set of constraints in \eqref{eq:IP}.

Given a fractional (or integral) solution $y$ and a subset of
sites $A\subseteq V$, let
$y_\ell(A)=\sum_{j\in A}y_{j\ell}$
denote the total mass of resource $\ell$ in $A$.
With a slight abuse of notation, we define
\begin{equation}\label{eq:cost}
C(y):=\sum_{i,\ell}w_{i\ell}\int_0^D\pos{1-y_\ell(B(i,r))}\,dr,
 \qquad F(y):=\sum_{j,\ell}f_{j\ell}y_{j\ell},
\end{equation}
where $(\cdot)_+=\max\{0,\cdot\}$. We note that if $y$ is the binary characteristic vector of a
feasible solution $S$, i.e., $y_{j\ell}=1$ if $j\in S_\ell$
and $y_{j\ell}=0$ otherwise, then $C(y)=C(S)$ and
$F(y)=F(S)$. For any fixed fractional solution $y\in Y$, the minimum
transportation cost in the fractional relaxation of
\eqref{eq:IP} is exactly $C(y)$. To see this, consider
a client-resource pair $(i,\ell)$. The optimal fractional
assignment allocates one unit of demand to sites in
nondecreasing order of their distance from $i$, with
each site $j$ receiving at most $y_{j\ell}$ units.
Consequently, the fraction of demand that remains
unassigned within distance $r$ is
$\pos{1-y_\ell(B(i,r))}$.
Integrating this quantity over $r\in[0,D]$ and summing
over all client-resource pairs, weighted by their demands,
gives exactly $C(y)$ as defined in \eqref{eq:cost}.
Therefore, the fractional relaxation of the data
placement problem can be written as
\begin{align}\label{eq:fractional-relax}
\operatorname{LP}=\min_{y \in Y} \{C(y)+F(y)\}.
\end{align}
This relaxation can be expressed as a polynomial-size linear
program by introducing an auxiliary variable for each positive-part
term, since the ball $B(i,r)$ changes only when $r$ crosses
an input distance in $\{d(i,j):i,j\in V\}$.
Moreover, as established above, when $y\in Y$ is integral,
the costs in \eqref{eq:cost} coincide with the actual
access and installation costs of the corresponding
feasible placement $S$. Therefore,
\eqref{eq:fractional-relax} is a valid relaxation of
the data placement problem, and hence
$\operatorname{LP}\le\OPT$.

\section{Multiscale Clustering and Flow Network Construction}\label{sec:balls}

In this section, we develop the main geometric and network
structures underlying our approximation algorithm. Starting
from an optimal solution $y$ of the fractional relaxation
\eqref{eq:fractional-relax}, we construct multiscale partitions
for each resource by greedily selecting disjoint metric balls
according to their $y_\ell$-mass. We then connect overlapping
balls at consecutive levels to form a directed flow network,
where the different resource layers are coupled through
the capacity constraints of the physical sites. An important
feature of this construction is that the fractional solution
$y$ naturally induces a feasible circulation on the network,
which will serve as a reference for the subsequent rounding
analysis. This network provides the foundation for the
surrogate-cost optimization developed in the next section.

\subsection{Multiscale Greedy Ball Partitions}\label{sec:multi-scale-partition}

Fix a fractional placement $y\in Y$ and radii $0<r_0<\cdots<r_L=D$. For each resource $\ell\in \mathcal{R}$ and level $s\in \{0,1,\ldots,L\}$, list every distinct ball centered at a site whose radius is zero or an input distance at most $r_s$. Scan them in decreasing order of $y_\ell$-mass, breaking ties in favor of larger balls, and retain a ball if it is disjoint from those already retained. Denote the retained family of balls by $\calP_{\ell,s}$. In particular, $\calP_{\ell,L}=\{V\}$. We note that the partitions at successive levels need not be nested, i.e., the balls in $\calP_{\ell,s}$ and $\calP_{\ell,s'}$ for $s'\neq s$ can overlap in any arbitrary fashion, and the construction is carried out separately for each resource $\ell$.

The above multiscale greedy clustering construction is motivated
by the integral representation of the access cost in
\eqref{eq:cost}, which measures the missing fractional mass
of each resource within metric balls of increasing radii.
Our construction captures this structure by considering
metric balls at different distance scales. To round a fractional placement into an integral one, we aim
to preserve this local resource mass as much as possible while
respecting the site-capacity constraints. However, directly
preserving the mass of every metric ball would introduce a
large collection of overlapping constraints. We therefore
consider a sequence of distance scales and construct a
disjoint family of representative balls for each resource
at every scale. By selecting balls greedily in decreasing
order of their $y_\ell$-mass, we ensure that every ball of
radius at most $r_s$ intersects a selected ball in
$\calP_{\ell,s}$ with at least as much fractional mass.
In this way, the resulting partitions capture the local
distribution of each resource across different distance scales.

\begin{lemma}\label{lem:balls}
Each $\calP_{\ell,s}$ partitions $V$ into balls of diameter at most $2r_s$. If $0\le r\le r_s$, then for every $i$ there is a ball $A\in\calP_{\ell,s}$ with
\begin{equation}\label{eq:greedy}
 A\cap B(i,r)\ne\varnothing,
 \qquad y_\ell(A)\ge y_\ell(B(i,r)).
\end{equation}
\end{lemma}
\begin{proof}
By the greedy selection rule, the balls in $\calP_{\ell,s}$ are pairwise disjoint, while the singleton candidates ensure that they cover $V$. Thus, each $\calP_{\ell,s}$ forms a partition of $V$. Now, fix any $0\le r\le r_s$. Since $B(i,r)$ coincides with a candidate ball of radius at most $r_s$, it was either selected by the greedy process, in which case we may take $A=B(i,r)$, or rejected because it intersected an earlier selected ball $A\in\calP_{\ell,s}$ whose $y_\ell$-mass is at least that of $B(i,r)$. In either case, \eqref{eq:greedy} follows.
\end{proof}

\subsection{Network Flow Construction}\label{sec:network}

The construction of our network is motivated by the fact that
the access and installation costs in \eqref{eq:cost} are separable
across resources, whereas the site-capacity constraints couple
their placement decisions. This suggests treating each resource
separately while enforcing the coupling through a single flow
network. To this end, we connect the multiscale greedy ball
partitions of each resource through directed arcs between
overlapping balls at consecutive levels. This structure allows
any positive integral flow through a ball to be traced backward
to an installed copy of the corresponding resource, with the
access distance bounded in terms of the radii along the flow
path. By coupling all resource layers through
shared site-capacity arcs, we obtain a single circulation
network that enforces the original capacity constraints
while allowing us to control the resulting access and
installation costs.

\begin{remark}
Another motivation for resource-wise decomposition comes from
Lagrangian relaxation of the site-capacity constraints, which
decomposes the problem into uncapacitated facility-location
problems with modified installation costs \cite{Etesami2024}.
However, this relaxation does not guarantee a feasible joint
placement, whereas our circulation couples the resource layers
while preserving feasibility.
\end{remark}

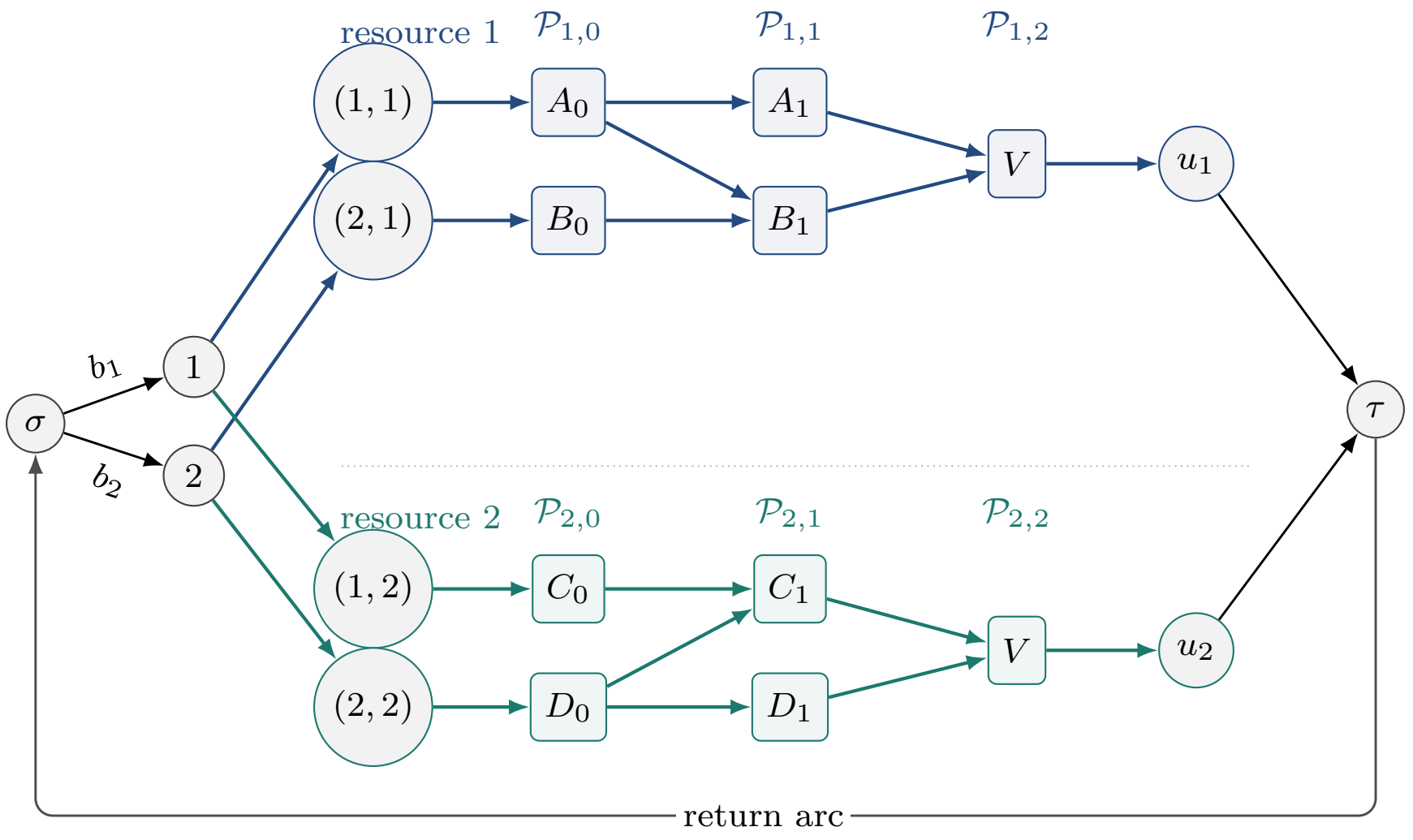
\begin{figure}[H]
\centering
\begin{tikzpicture}[x=.86cm,y=.77cm,>=Latex,every node/.style={font=\scriptsize}]
  \node[flowgate] (sig) at (0,-.8) {$\sigma$};
  \node[flowgate] (p1) at (1.5,-.2) {$1$};
  \node[flowgate] (p2) at (1.5,-1.35) {$2$};
  \draw[flowarrow] (sig) -- (p1) node[pos=.52,above,sloped] {$b_1$};
  \draw[flowarrow] (sig) -- (p2) node[pos=.55,below,sloped] {$b_2$};
  \node[flowgate,draw=navy] (q11) at (3.2,2.6) {$(1,1)$};
  \node[flowgate,draw=navy] (q21) at (3.2,1.35) {$(2,1)$};
  \node[flowgate,draw=teal] (q12) at (3.2,-2.55) {$(1,2)$};
  \node[flowgate,draw=teal] (q22) at (3.2,-3.8) {$(2,2)$};
  \draw[primary] (p1) -- (q11);
  \draw[primary] (p2) -- (q21);
  \draw[secondary] (p1) -- (q12);
  \draw[secondary] (p2) -- (q22);
  \node[flowball] (a0) at (5.05,2.6) {$A_0$};
  \node[flowball] (b0) at (5.05,1.35) {$B_0$};
  \node[flowball] (a1) at (7.15,2.6) {$A_1$};
  \node[flowball] (b1) at (7.15,1.35) {$B_1$};
  \node[flowball] (v1) at (9.3,1.95) {$V$};
  \node[flowgate,draw=navy] (u1) at (11,1.95) {$u_1$};
  \draw[primary] (q11) -- (a0); \draw[primary] (q21) -- (b0);
  \draw[primary] (a0) -- (a1);
  \draw[primary] (b0) -- (b1);
  \draw[primary] (a0) -- (b1);
  \draw[primary] (a1) -- (v1); \draw[primary] (b1) -- (v1);
  \draw[primary] (v1) -- (u1);
  \node[flowballtwo] (c0) at (5.05,-2.55) {$C_0$};
  \node[flowballtwo] (d0) at (5.05,-3.8) {$D_0$};
  \node[flowballtwo] (c1) at (7.15,-2.55) {$C_1$};
  \node[flowballtwo] (d1) at (7.15,-3.8) {$D_1$};
  \node[flowballtwo] (v2) at (9.3,-3.2) {$V$};
  \node[flowgate,draw=teal] (u2) at (11,-3.2) {$u_2$};
  \draw[secondary] (q12) -- (c0); \draw[secondary] (q22) -- (d0);
  \draw[secondary] (c0) -- (c1); \draw[secondary] (d0) -- (d1);
  \draw[secondary] (d0) -- (c1);
  \draw[secondary] (c1) -- (v2); \draw[secondary] (d1) -- (v2);
  \draw[secondary] (v2) -- (u2);
  \node[flowgate] (tau) at (12.7,-.65) {$\tau$};
  \draw[flowarrow] (u1) -- (tau); \draw[flowarrow] (u2) -- (tau);
  \draw[flowarrow,rounded corners=4pt,black!70] (tau.south) |- (7,-4.95) -| (sig.south);
  \node[fill=white,inner sep=1.5pt] at (6.9,-4.95) {return arc};
  \foreach \xx/\ss in {5.05/0,7.15/1,9.3/2}{
    \node[navy] at (\xx,3.39) {$\mathcal P_{1,\ss}$};
    \node[teal] at (\xx,-1.77) {$\mathcal P_{2,\ss}$};
  }
  \node[navy] at (3.65,3.35) {resource $1$};
  \node[teal] at (3.65,-1.79) {resource $2$};
  \draw[black!25,densely dotted] (2.9,-1.25) -- (11.5,-1.25);
\end{tikzpicture}
\caption{One globally coupled network for two resources $\mathcal{R}=\{1,2\}$, two sites $V=\{1,2\}$, and three
ball levels $s=0,1,2$. Physical sites $1$ and $2$ share the site-capacity arcs,
while $(j,\ell)$ denotes the copy of site $j$ in resource layer $\ell$. The interlevel arcs illustrate set-theoretic overlaps.}
\label{fig:network}
\end{figure}

Given a fractional solution $y$, construct and fix the
greedy ball partitions $\calP_{\ell,s}$. We then build
a directed network whose structure is determined by the
physical sites and the \emph{set-theoretic overlaps}
of the selected balls, as illustrated in
Figure~\ref{fig:network}. The network is constructed
as follows:\footnote{All arcs without explicitly specified capacity bounds in the
following network construction have lower bound zero and
upper bound $n$.}
\begin{itemize}
    \item Introduce a source node $\sigma$, a sink node $\tau$,
    and a return arc $\tau\to\sigma$ with capacity $nm$.
    \item For each physical site $j\in V$, introduce a node $j$
    and an arc $\sigma\to j$ with capacity $b_j$. These are the
    only arcs that couple different resources.
    \item For every site-resource pair in $V\!\times\!\R$, introduce a node $(j,\ell)$ and an arc
    $j\!\to\!(j,\ell)$ with capacity 1.
    \item For each resource $\ell$, level $s$, and selected ball
    $A\in\calP_{\ell,s}$, introduce one node $A$. Different resource-level nodes remain distinct even when
    they represent the same subset of $V$. For every $(j,\ell)$,
    add an arc to the unique first-level ball node
    $A\in \calP_{\ell,0}$ with $j\in A$. For each $0\le s<L$, add an arc
    $A\to B$ whenever they intersect, i.e., $A\in\calP_{\ell,s}, B\in\calP_{\ell,s+1}$, $A\cap B\ne\varnothing$.
    \item For each resource $\ell$, connect its final-level ball
    node $\calP_{\ell,L}=\{V\}$ to a \emph{root} node $u_\ell$, and add the arc
    $u_\ell\to\tau$ with upper-capacity $n$ and lower-capacity 1. The lower bounds on the root capacities are used solely to ensure that at least one copy of each resource is installed in the network.  
\end{itemize}

\subsection{Reference Circulation}
We now use the fractional placement $y$ to construct a
feasible reference circulation on the network described above,
which will serve as a benchmark for the rounding analysis
in the next section. For each resource $\ell$, site $j$,
and level $s$, let $A_{\ell,s}(j)$ denote the unique ball
in $\calP_{\ell,s}$ containing $j$. We route $y_{j\ell}$
units of flow along the directed cycle
\[
\sigma\longrightarrow j\longrightarrow(j,\ell)
\longrightarrow A_{\ell,0}(j)\longrightarrow A_{\ell,1}(j)\longrightarrow\cdots
\longrightarrow A_{\ell,L}(j)\longrightarrow u_\ell
\longrightarrow\tau\longrightarrow\sigma.
\]
Every arc on this cycle belongs to the constructed network,
since consecutive balls contain $j$ and therefore intersect.
Summing these cycle flows over all site-resource pairs
$(j,\ell)$ defines a fractional circulation $\tilde{\varphi}$.
For each arc $a\to b$, let $\tilde{\varphi}(a\to b)$ denote
the resulting flow on that arc. By construction,
\begin{equation}\label{eq:reference}
\begin{aligned}
 \tilde{\varphi}(\sigma\to j)&=\sum_{\ell\in\R}y_{j\ell},\qquad
 &\tilde{\varphi}(j\to(j,\ell))&=y_{j\ell},\\
 \tilde{\varphi}\big((j,\ell)\to A_{\ell,0}(j)\big)&=y_{j\ell},
 &\tilde{\varphi}(A\to B)&=y_\ell(A\cap B),\\
 \tilde{\varphi}(u_\ell\to\tau)&=y_\ell(V),
 &\tilde{\varphi}(\tau\to\sigma)&=\sum_{j\in V}\sum_{\ell\in\R}y_{j\ell},
\end{aligned}
\end{equation}
where $A\to B$ denotes an arc from $A\in\calP_{\ell,s}$
to $B\in\calP_{\ell,s+1}$. Flow conservation holds because
$\tilde{\varphi}$ is a sum of flows along directed cycles.
Moreover, all capacity constraints are satisfied since
$y\in Y$, and the lower bound at each resource root is
satisfied because $y_\ell(V)\ge1$. Thus,
$\tilde{\varphi}$ is a feasible circulation that is supported on the arcs of the constructed network.

\section{Rounding}
\label{sec:rounding}
\renewcommand{\theHequation}{\thesection.\arabic{equation}}

In this section, we show how to round the reference circulation $\tilde{\varphi}$ to an integral one with a provable performance guarantee. To this end, we introduce the following definition, which
serves as a flow-based proxy for the fractional mass of a ball.

\begin{definition}
For any feasible circulation $\varphi$ on the constructed network and any ball node $A\in\calP_{\ell,s}$, let $v_A(\varphi)$ denote the throughput of $A$, defined as the sum of its incoming flows or, its outgoing flows.
\end{definition}

\begin{remark}\label{rem:throughput}
For the reference circulation $\tilde{\varphi}$, since the
balls at each level partition $V$, it follows from
\eqref{eq:reference} that
$v_A(\tilde{\varphi})=y_\ell(A)$ for any ball node $A\in\calP_{\ell,s}$.
\end{remark}

Every integral feasible circulation $\varphi$ on the network defines a placement
\begin{align}\label{eq:placement-S}
S_\ell(\varphi):=\{j\in V:\varphi(j\to(j,\ell))=1\}.
\end{align}
This placement is feasible: the site-capacity arcs enforce $\sum_{\ell\in\R}\mathbf{1}\{j\in S_\ell(\varphi)\}\le b_j$, and the root lower bounds ensure that every resource has at least one copy. To bound the cost of this placement in terms of $\OPT$, we must round the reference circulation while controlling the resulting access and placement costs. The following lemma provides the geometric connection needed to control access costs: positive integral flow through a ball guarantees an installed copy of the corresponding resource near that ball.

\begin{lemma}\label{lem:geometry}
Let $A\in\calP_{\ell,s}$. If $\varphi$ is an \emph{integral} feasible circulation and $v_A(\varphi)\ge1$, then some installed copy of $\ell$ lies within distance $2\sum_{k=0}^{s-1}r_k$ of $A$.\footnote{Here, the sum is understood to be zero when $s=0$.}Consequently, if $A\cap B(i,r)\ne\varnothing$ and $r\le r_s$, then
\begin{equation}\nonumber
v_A(\varphi)\ge1
\quad\Longrightarrow\quad
d(i,S_\ell(\varphi))\le r+2\sum_{k=0}^{s}r_k.
\end{equation}
\end{lemma}

\begin{figure}[!htbp]
\centering
\begin{tikzpicture}[x=1cm,y=1cm,>=Latex,font=\small]
  \fill[navy!7] (1.55,.10) circle (1.22);
  \fill[teal!7] (3.45,.10) circle (1.26);
  \fill[navy!7] (5.35,.10) circle (1.26);
  \draw[navy,thick] (1.55,.10) circle (1.22);
  \draw[teal,thick] (3.45,.10) circle (1.26);
  \draw[navy,thick] (5.35,.10) circle (1.26);
  \draw[gold,thick,dashed] (7.22,.10) circle (.98);
  \coordinate (j) at (.70,.10);
  \coordinate (k0) at (2.52,.10);
  \coordinate (k1) at (4.39,.10);
  \coordinate (k2) at (6.35,.10);
  \coordinate (i) at (7.22,.10);
  \draw[flowarrow,black!70] (j) -- (k0) -- (k1) -- (k2) -- (i);
  \foreach \z in {j,k0,k1,k2,i}{\fill[black!80] (\z) circle (2pt);}
  \node[below] at (j) {$j\in S_\ell(\varphi)$};
  \node[below right] at (i) {$i$};
  \node[navy,above] at (1.45,1.44) {$A_0$};
  \node[teal,above] at (3.45,1.50) {$A_1$};
  \node[navy,above] at (5.35,1.50) {$A_2=A$};
  \node[gold,above right] at (7.78,.7) {$B(i,r)$};
  \node[font=\scriptsize] at (1.55,-.55) {$\le 2r_0$};
  \node[font=\scriptsize] at (3.45,-.55) {$\le 2r_1$};
  \node[font=\scriptsize] at (5.35,-.55) {$\le 2r_2$};
  \node[font=\scriptsize] at (6.87,-.55) {$\le r$};
  \node[font=\small,align=center] at (4,-1.55)
   {$d(i,j)\le r+2(r_0+r_1+r_2)$};
\end{tikzpicture}
\caption{Lemma~\ref{lem:geometry} for three levels: a unit of integral flow through $A_2$ can be traced back to an installed site $j$ through a sequence of consecutively intersecting balls. The sum of their diameter bounds and the radius $r$ bounds the distance from $i$ to $j$.}
\label{fig:geometry}
\end{figure}
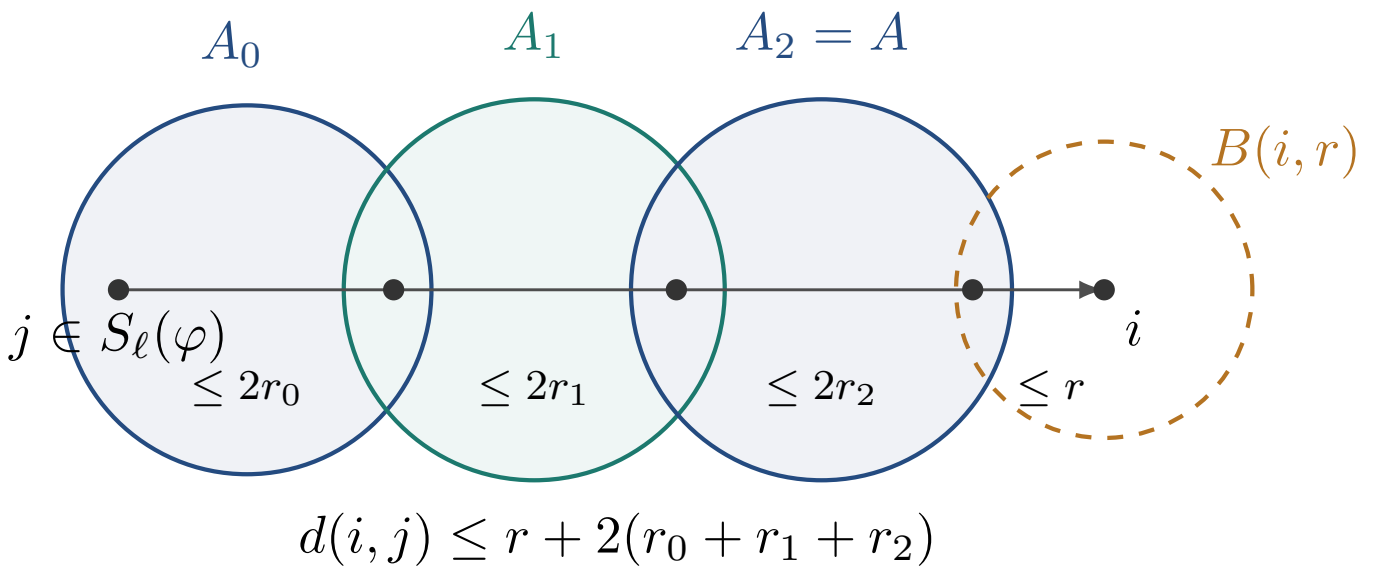

\begin{proof}
Trace a unit of flow backward from $A$ through the preceding levels to a resource-site entry $(j,\ell)$. Since the arc $j\to(j,\ell)$ has unit capacity and carries positive integral flow, $j\in S_\ell(\varphi)$. The balls visited along this path, denoted by $A_0,\ldots,A_s=A$, intersect consecutively and have diameters at most $2r_0,\ldots,2r_s$, respectively (see Figure \ref{fig:geometry}). Since $j\in A_0$, these intersections imply $d(j,A)\le2\sum_{k=0}^{s-1}r_k$.
Moreover, for any $a\in A\cap B(i,r)$ with $r\leq r_s$, the same chain of intersections gives $d(j,a)\le2\sum_{k=0}^{s}r_k$.
Since $d(i,a)\le r$, the triangle inequality yields
\[
d(i,S_\ell(\varphi))
\le d(i,j)
\le d(i,a)+d(a,j)
\le r+2\sum_{k=0}^{s}r_k.
\]
\end{proof}

\vspace{-0.3cm}

\subsection{Rounding Using a Surrogate Flow Cost}

We now round the reference circulation by charging for missing flow near each client. Recall that the
original fractional cost is
\begin{equation}\label{eq:p}
F(y)+C(y)
=
\sum_{i,\ell}f_{i\ell}y_{i\ell}
+
\sum_{i,\ell}w_{i\ell}
\int_0^D \pos{1-y_\ell(B(i,r))}\,dr.
\end{equation}

Our goal is to construct a surrogate cost function $Q(\cdot)$
as a function of the flow over the constructed network,
with two properties. First, $Q$ should
upper-bound the actual installation and access costs
of every integral circulation. Second, its value at
the reference circulation $\tilde\varphi$ should be
upper-bounded by the fractional cost \eqref{eq:p}
plus a controllable correction term. These properties
will allow us to minimize $Q$ over the flow network,
which we will show admits an integral optimal circulation,
and compare the resulting integral placement directly
with the optimal value of the fractional LP.

To achieve this, we seek to preserve the structure of \eqref{eq:p} by replacing the fractional
mass of each ball with the flow throughput of a
corresponding network witness. Here, a witness is a
selected ball in the constructed network whose
positive integral flow certifies that a particular
resource is installed sufficiently close to a client. To construct these witnesses, set $r_{-1}=0$ and
$h_s=2\sum_{k=0}^s r_k$. For each demand $(i,\ell)$
and radius $r\in(0,D]$, we select a network witness
$A_{i\ell}(r)$ to represent the ball $B(i,r)$
appearing in \eqref{eq:p}. Specifically, letting
$s$ be the first level with $r\le r_s$,
Lemma~\ref{lem:balls} allows us to select
$A_{i\ell}(r)\in\calP_{\ell,s}$ satisfying\footnote{Note that these witnesses are fixed using only the fractional
solution $y$ and the prescribed radii, independently
of the circulation to be optimized.}
\begin{equation}\label{eq:witness}
A_{i\ell}(r)\cap B(i,r)\ne\varnothing,
\qquad
y_\ell(A_{i\ell}(r))
\ge y_\ell(B(i,r)).
\end{equation}
In particular,
at the reference circulation $\tilde\varphi$, by Remark \ref{rem:throughput} we have
\[
v_{A_{i\ell}(r)}(\tilde\varphi)
=y_\ell(A_{i\ell}(r))
\ge y_\ell(B(i,r)).
\]
Since $(1-x)_+$ is nonincreasing in $x$,
replacing the fractional ball masses in \eqref{eq:p} by their witness
throughputs can only decrease the access-cost
integral at the reference circulation $\tilde{\varphi}$, as desired.

However, this replacement alone does not guarantee an upper bound
on the actual cost of an integral circulation $\varphi$.
Indeed, by Lemma~\ref{lem:geometry}, positive integral flow
through $A_{i\ell}(r)$, for $r\in(r_{s-1},r_s]$,
certifies coverage within distance $r+h_s$, but not
necessarily within distance $r$. We must therefore account
for the additional distance introduced by clustering. To this end, we sweep the radius $r$ over the integration
interval $[0,D]$, replace the fractional mass
$y_{\ell}(B(i,r))$ with the corresponding witness throughput
$v_{A_{i\ell}(r)}(\varphi)$ defined in \eqref{eq:witness},
and separately charge for the additional geometric
expansion $h_s$ at each level. 

At the first level, the coverage guarantee incurs an initial
expansion $h_0=2r_0$. We charge this distance
whenever the entry arc $i\to(i,\ell)$ carries
no flow, since positive flow through this arc
certifies that resource $\ell$ is stored at
site $i$ itself. At each subsequent level $s$, the additional
expansion relative to the preceding level is
$h_s-h_{s-1}=2r_s$. We charge this increment
whenever the preceding witness
$A_{i\ell}(r_{s-1})$ carries no flow.
Indeed, positive integral flow through this
witness already certifies coverage within
distance $r_{s-1}+h_{s-1}$, so no additional
charge is needed for the intervening distance.
Otherwise, charging the full increment
accounts for the possible delay before the
coverage guarantees of level $s$ become valid. Consequently, for each demand $(i,\ell)$,
the surrogate access cost consists of the
original-radius integral of missing witness
flow, together with the geometric correction
\[
\int_0^D
\pos{1-v_{A_{i\ell}(r)}(\varphi)}\,dr+
h_0\pos{1-\varphi(i\to(i,\ell))}+
\sum_{s=1}^L(h_s-h_{s-1})
\pos{1-v_{A_{i\ell}(r_{s-1})}(\varphi)}.
\]
For convenience, set
$A_{i\ell}(0):=i\to(i,\ell)$, with
$v_{A_{i\ell}(0)}(\varphi)
=\varphi(i\to(i,\ell))$.
Using $h_0=2r_0$ and
$h_s-h_{s-1}=2r_s$, and adding the
installation cost in terms of flow, we obtain the surrogate cost
\begin{align}\label{eq:flow-cost}
Q(\varphi)
=\sum_{j,\ell}f_{j\ell}
\varphi(j\to(j,\ell))+\sum_{i,\ell}w_{i\ell}\Bigg[
\int_0^D
\pos{1-v_{A_{i\ell}(r)}(\varphi)}\,dr+2\sum_{s=0}^L r_s
\pos{1-v_{A_{i\ell}(r_{s-1})}(\varphi)}
\Bigg].
\end{align}


\begin{lemma}\label{lem:flow-cost}
For every integral feasible circulation $\varphi$,
the induced placement $S(\varphi)$ defined by \eqref{eq:placement-S} satisfies
\[
C(S(\varphi))+F(S(\varphi))\le Q(\varphi).
\]
Moreover, for the reference circulation $\tilde\varphi$,
\[
Q(\tilde\varphi)
\le
F(y)+C(y)+2\sum_{s=0}^L r_s p(r_{s-1}),
\]
where $p(t):=\sum_{i,\ell}w_{i\ell}
\pos{1-y_\ell(B(i,t))}$. Finally, minimizing $Q$ over the feasible circulation
polytope admits an integral optimal circulation,
which can be found by a min-cost circulation algorithm.
\end{lemma}

\begin{proof}
We first show that $Q$ upper-bounds the cost of any
integral feasible circulation $\varphi$.
Fix a demand $(i,\ell)$. If the entry arc
$i\to(i,\ell)$ carries positive flow, then
$i\in S_\ell(\varphi)$ and the access distance is zero.
Otherwise, let $r\in(r_{s-1},r_s]$ be the first radius
at which $A_{i\ell}(r)$ carries positive flow
(taking a limit if necessary).
Since all earlier witnesses have zero flow,
the missing-flow integral in \eqref{eq:flow-cost}
contributes at least $r$, while the correction terms
through level $s$ contribute at least
$2\sum_{k=0}^s r_k=h_s$.
By Lemma~\ref{lem:geometry}, the positive flow
certifies an installed copy within distance $r+h_s$.
Thus, the total charge is at least the actual access
distance $d(i,S_\ell(\varphi))$. If no witness carries positive flow, the missing-flow
integral contributes $D$, which also bounds the
access distance by feasibility.
Multiplying by $w_{i\ell}$, summing over all demands,
and using $F(S(\varphi))
=\sum_{j,\ell}f_{j\ell}\varphi(j\to(j,\ell))$ gives $C(S(\varphi))+F(S(\varphi))\le Q(\varphi)$, as claimed.

Next, consider the reference circulation
$\tilde\varphi$. By Remark \ref{rem:throughput}, each witness
throughput equals its fractional mass. Thus,
\eqref{eq:witness} implies
\[
\pos{1-v_{A_{i\ell}(r)}(\tilde\varphi)}
\le
\pos{1-y_\ell(B(i,r))}.
\]
The same inequality holds at $r=0$, since
$A_{i\ell}(0)$ is the entry arc and
$B(i,0)=\{i\}$. Applying these inequalities
to \eqref{eq:flow-cost} gives
\[
\begin{aligned}
Q(\tilde\varphi)
\le F(y)+\int_0^D p(r)\,dr
+2\sum_{s=0}^L r_s p(r_{s-1})=F(y)+C(y)
+2\sum_{s=0}^L r_s p(r_{s-1}),
\end{aligned}
\]
proving the second claim.

Finally, we establish integrality of an optimal
circulation. Since the network contains finitely
many witnesses, the integral and correction terms
in \eqref{eq:flow-cost} can be grouped by their
corresponding node throughputs. Thus, $Q$ is a
sum of linear installation costs and nonnegative
weighted terms of the form
$c_A\pos{1-v_A(\varphi)}$, where $c_A\ge 0$.
Each such term is convex and piecewise linear
in the node throughput, with marginal cost
$-c_A$ on $[0,1]$ and zero beyond $1$. To represent this cost linearly, we split each
witness node $A$ into an entrance and an exit,
redirecting all incoming arcs to the entrance
and all outgoing arcs from the exit. We then
connect these two nodes by two parallel arcs:
one with capacity $1$ and unit cost $-c_A$,
and the other with zero unit cost and unlimited
capacity. Since the first arc has no greater
cost, an optimal circulation can be chosen to send
$\min\{1,v_A(\varphi)\}$ units through it.
The resulting cost is $-c_A\min\{1,v_A(\varphi)\}
=c_A\pos{1-v_A(\varphi)}-c_A$, which recovers the desired term up to an additive constant. Consequently, minimizing $Q$ is equivalent,
up to an additive constant independent of
$\varphi$, to a linear min-cost circulation
problem on the expanded network. Since all
capacities and lower bounds remain integral,
the network-flow integrality theorem guarantees
an integral optimal circulation. Merging the split nodes then yields an integral feasible
circulation that minimizes $Q$ over the original network
constructed in Section~\ref{sec:network}.
\end{proof}

\section{Optimizing the Radii and Approximation Guarantee}
\label{sec:scales}

By Lemma~\ref{lem:flow-cost}, the approximation guarantee
depends on the choice of radii through the correction term
$\sum_{s=0}^L r_s p(r_{s-1})$. We show that these radii
can be chosen in polynomial time so that this term is
at most $eC(y)$, leading to our main approximation
guarantee. The complete procedure is summarized in
Algorithm~\ref{alg:placement}, which combines the
fractional relaxation, multiscale greedy clustering,
and surrogate flow optimization into a deterministic
rounding algorithm. The following theorem establishes
its performance guarantee.

\begin{algorithm}[t]
\caption{Multiscale Flow Rounding for Metric Data Placement}
\label{alg:placement}
\begin{algorithmic}[1]
\Require Metric instance $(V,d)$, resources $\R$,
capacities $b_j$, installation costs $f_{j\ell}$,
and demands $w_{i\ell}$.

\State Solve the fractional relaxation
\eqref{eq:fractional-relax} to obtain an optimal $y\in Y$.

\State Compute
$p(r)=\sum_{i,\ell}w_{i\ell}
\pos{1-y_\ell(B(i,r))}$.
Choose radii $0<r_0<\cdots<r_L=D$ by minimizing
$\sum_{s=0}^L r_s p(r_{s-1})$
over the shifted geometric sequences (Theorem \ref{thm:main}).

\State For every resource $\ell$ and level $s$,
construct the greedy ball partition $\calP_{\ell,s}$
as in Section~\ref{sec:multi-scale-partition}.

\State Construct the circulation
network using the selected balls and their overlaps,
as in Section~\ref{sec:network}.

\State Select the witnesses $A_{i\ell}(r)$ according
to \eqref{eq:witness} and define the surrogate
flow cost $Q(\varphi)$ in \eqref{eq:flow-cost}.

\State Compute an integral optimal circulation
$\varphi^*$ minimizing $Q(\varphi)$ using the
min-cost flow (Lemma~\ref{lem:flow-cost}).

\State \Return $S=(S_\ell)_{\ell\in\R}$, where
$S_\ell=\{j\in V:
\varphi^*(j\to(j,\ell))=1\}$.

\end{algorithmic}
\end{algorithm}

\begin{theorem}\label{thm:main}
Algorithm~\ref{alg:placement} runs in deterministic
polynomial time and returns a feasible placement
$S$ satisfying
\[
C(S)+F(S)
\le F(y)+(1+2e)C(y)
\le (1+2e)\OPT,
\]
where $y$ is an optimal solution of the fractional
relaxation \eqref{eq:fractional-relax}.
\end{theorem}

\begin{proof}
We first show that there exist radii $0<r_0<\cdots<r_L=D$ satisfying
\begin{equation}\label{eq:e}
\sum_{s=0}^L r_s p(r_{s-1})\le eC(y),
\end{equation}
where we recall that $p(t):=\sum_{i,\ell}w_{i\ell}
\pos{1-y_\ell(B(i,t))}$. To this end, let $0=t_0<t_1<\cdots<t_M=D$ be the distinct input distances. Since $p$ is a nonincreasing step function that changes
only at the input distances $t_1,\ldots,t_M$, we have
$p(t)=p(t_{k-1})$ for $t\in[t_{k-1},t_k)$.
Moreover, since $p(D)=0$, we have
\begin{align}\label{eq:C-p}
C(y)=\int_0^D p(t)\,dt
=\sum_{k=1}^M p(t_{k-1})(t_k-t_{k-1})
=\sum_{k=1}^M \big(p(t_{k-1})-p(t_k)\big)t_k.
\end{align}
Similarly, for any chosen radii,
\begin{align}\label{eq:r-p-sum}
\sum_{s=0}^L r_s p(r_{s-1})
=\sum_{s=0}^L r_s
\sum_{k:\,t_k>r_{s-1}}
\bigl(p(t_{k-1})-p(t_k)\bigr)
=\sum_{k=1}^M
\bigl(p(t_{k-1})-p(t_k)\bigr)
\sum_{\substack{0\leq s\leq L:\\ r_{s-1}<t_k}}r_s.
\end{align}

Now, let $u\in[0,1)$ and consider the geometric sequence of
radii $\delta e^{s+u}$, $s=0,1,2,\ldots$.
Round each radius down to the largest input distance
not exceeding it, discard repeated values, and terminate
the sequence at $D$. Denote the resulting rounded radii sequence by
$0<r_0(u)<\cdots<r_{L(u)}(u)=D$, where $r_{-1}(u):=0$. Evaluating \eqref{eq:r-p-sum} at these rounded radii and integrating over $u\in[0,1)$ gives
\[
\begin{aligned}
\int_0^1 \sum_{s=0}^{L(u)} r_s(u) p(r_{s-1}(u))\,du
&=\sum_{k=1}^M
\bigl(p(t_{k-1})-p(t_k)\bigr)
\int_0^1\sum_{\substack{0\le s\le L(u):\\ \,r_{s-1}(u)<t_k}}r_s(u)\,du\\
&\le \sum_{k=1}^M
\bigl(p(t_{k-1})-p(t_k)\bigr)
\int_0^1
\sum_{\substack{s\ge0:\\ \delta e^{s-1+u}<t_k}}
\delta e^{s+u}\,du\\
&=\sum_{k=1}^M
\bigl(p(t_{k-1})-p(t_k)\bigr)
\Big((e-1)\delta+e(t_k-\delta)\Big)\\
&\le e\sum_{k=1}^M
\bigl(p(t_{k-1})-p(t_k)\bigr)t_k
=eC(y).
\end{aligned}
\]
where the last equality holds by \eqref{eq:C-p}. Therefore, there exists $u^*\in[0,1)$ such that the corresponding radii sequence $r_s:=r_s(u^*)$, $0\le s\le L(u^*)$, satisfies \eqref{eq:e}. In fact, such a choice can be found
deterministically: the rounded radii sequence $0<r_0(u)<\cdots<r_{L(u)}(u)=D$ changes only
when $\delta e^{s+u}=t_k$ for some input distance
$t_k$. Each $t_k$ determines at most one such
$u\in[0,1)$, so $u^*$ can be found by examining these at most $M$
critical shifts and $u=0$ and selecting the one that minimizes the expression in \eqref{eq:r-p-sum}.

Finally, construct the flow network using these
radii and minimize the surrogate cost $Q$ in
\eqref{eq:flow-cost}. By Lemma~\ref{lem:flow-cost},
an integral optimal circulation $\varphi^*$ exists,
and its induced placement $S:=S(\varphi^*)$ satisfies
\[
\begin{aligned}
C(S)+F(S)
&\le Q(\varphi^*)
\le Q(\tilde\varphi)\\
&\le F(y)+C(y)
   +2\sum_{s=0}^L r_s p(r_{s-1})\\
&\le F(y)+(1+2e)C(y)
\le (1+2e)\OPT,
\end{aligned}
\]
where the last inequality follows from
$F(y)\ge0$ and
$F(y)+C(y)=\operatorname{LP}\le\OPT$.

The algorithm runs in polynomial time because there are at most
$M\le\binom{n}{2}$ distinct input distances and polynomially
many candidate radius sequences. For the selected sequence,
the greedy ball partitions and the expanded flow network
have polynomial size, and Lemma~\ref{lem:flow-cost} reduces
the rounding step to a single min-cost circulation problem.
This completes the proof.
\end{proof}

\section{Conclusion and Future Directions}
\label{sec:conclusion}

In this paper, we developed a deterministic polynomial-time
$(1+2e)$-approximation algorithm for the metric data placement
problem with heterogeneous demands, installation costs, and
integral site capacities, improving upon the previous
best-known approximation factor of $8$. Our approach builds
on a natural LP relaxation whose access cost admits a continuous
representation in terms of missing fractional resource mass
within metric balls. Using this representation, we constructed
multiscale greedy ball partitions independently for each
resource and connected overlapping balls at consecutive
levels to form a single flow network that captures the
coupling among resources through shared site-capacity constraints.
By minimizing a suitable surrogate cost over this network
and optimizing the choice of ball radii, we obtained an
integral placement whose total cost is at most
$(1+2e)\approx 6.44$ times the optimum. Beyond improving
the approximation guarantee, our results demonstrate how
the geometric structure of the objective function can be
exploited to develop a simple rounding procedure based
on a single min-cost circulation.

Several directions remain open for future research. One
particularly interesting question is whether the simple
combinatorial greedy algorithm developed in
\cite{Etesami2020}, which achieves a $3$-approximation
for homogeneous metric data placement, can be extended
to heterogeneous demands and installation costs while
retaining a constant-factor guarantee. Such an extension
could provide a purely combinatorial alternative to
LP-based rounding and potentially lead to more efficient
algorithms. Interestingly, the algorithm in
\cite{Etesami2020} also admits a greedy ball interpretation,
but employs a different online selection rule that accounts
for resource diversity. Another promising direction is to
investigate whether more refined multiscale clustering
schemes or richer network flow constructions can further
improve the approximation factor below $1+2e$ by
preserving more local information about resource
distributions across different distance scales.
Finally, it would be interesting to extend the proposed
framework to online and distributed settings, where
resource demands, storage capacities, or network
conditions may evolve over time, while maintaining
provable performance guarantees.

\paragraph{Use of AI.} The author developed the high-level approach, including the
integral ball-based relaxation, resource-wise decomposition,
and network flow construction using fractional ball partitions.
Initial explorations with ChatGPT (GPT-6) based on laminar
clustering yielded a preliminary algorithm without improving
the approximation factor. The author proposed a more general construction based on overlapping balls, leading to an improved approximation guarantee through iterative discussions with ChatGPT. Subsequently, the author substantially refined and simplified the construction, completed the missing parts and proofs, and used ChatGPT to improve the manuscript's editing and exposition.

\bibliographystyle{plain}
\bibliography{references}

\end{document}